\documentclass[aps,prb,twocolumn]{revtex4-1} 

\usepackage{amsmath,amssymb}

\let\origref\ref
\def\ref#1{\unskip~\origref{#1}}
\let\origcite\cite
\def\cite#1{\unskip~\origcite{#1}}
\newcommand{\refe}[1]{\unskip~(\origref{#1})}

\renewcommand{\vec}[1]{\mathbf{#1}}

\renewcommand{\r}{\vec{r}}
\newcommand{\Vb}{\bar{V}}

\newcommand{\bra}[1]{\langle #1 |}
\newcommand{\ket}[1]{| #1 \rangle}

\newcommand{\be}{\begin{equation}}
\newcommand{\bel}[1]{\begin{equation}\label{#1}}
\newcommand{\ee}{\end{equation}}
\newcommand{\bea}{\begin{eqnarray}}
\newcommand{\ea}{\end{eqnarray}}

\newcommand{\Schro}{Schr\"o\-dinger }

\begin{document}

\title{Comparing energy levels in isotropic and anisotropic potentials}
\author{Alexander Pikovski}
\email{E-mail: {\tt alexander.pikovski@colorado.edu}}
\affiliation{Department of Physics and JILA, University of Colorado, Boulder, CO 80309}
\date{\today}


\begin{abstract}
Qualitative information about the quantized energy levels of a system can be of great value. 
We study the relationship between the bound-state energies of an anisotropic potential and those of its spherical average. It is shown that the two ground-state energies satisfy an inequality, and there is a similar inequality for the first excited states. 
\end{abstract}


\maketitle


Two-body bound states can play an important role in physical properties of a many-body system; examples
are the thermodynamic equation of state of a gas \cite{Pais-Uhlenbeck},
or the energy of an impurity in an interacting gas \cite{Klawunn-Recati}.
The knowledge of general properties of bound states is useul both for concrete calculations as well as a help to find model potentials when the exact
interaction is unknown.

In this note, we derive general inequalities which relate the energies of quantum-mechanical bound states, the ground state and the first excited state, in an arbitrary 
anisotropic potential $V(\r)$ to the bound-state energies of the isotropic potential $\Vb$, which is the spherical average of $V$.
Inequalities of this type have not been discussed in connection with bound states, to the author's knowledge, although there are works in the mathematical literature \cite{Beesack,Banks} which consider a related ground-state inequality. 


Consider the bound states of one particle in an external anisotropic (non-central) potential $V(\r)$, described by the
\Schro equation
\bel{Schro}
H \psi = \Bigl\{ - \frac{\hbar^2}{2m}\nabla^2 + V(\r) \Bigr\} \psi(\r) = E \psi(\r).
\ee
We want to compare bound-state energies in a potential
$V(\r)$ with the bound-state energies in the spherically averaged potential. 
Given an arbitrary potential $V(\r)$, its spherical average is
\bel{sph-av}
\Vb (r) 
= \frac{1}{4\pi} \int_0^\pi \! \int_0^{2\pi} \!  V(r,\theta,\phi) \, \sin \theta d\theta d\phi 
\ee
where $(r,\theta,\phi)$ are spherical coordinates;
for short we write $d\Omega=\sin \theta d\theta d\phi$.
Clearly, the spherically averaged potential $\Vb$ itself is
spherically symmetric.
One can always write 
$V(\r) = \Vb(|\r|) + \Delta V (\r)$,
thus $\Vb$ may be considered as the isotropic part of $V$.
One has
\bel{dv0}
\int \! \Delta V(\r) \, d\Omega = 0
\ee
which follows directly from the definition, Eq. \refe{sph-av}.

\section{Ground state inequality}
\label{sec:gs}

Let us first derive a simple inequality which will be of use.
Consider, besides the Hamiltonian $H=H_0+V$, a second Hamiltonian
$H_U = H_0 + U$.
Let $u_0$ be the normalized wave function of the ground state of $H_U$ with energy $\epsilon_0 = \bra{u_0}H_U\ket{u_0}$. Let us {\em assume} that
\begin{align}\label{ass}
\bra{u_0} V - U \ket{u_0} \le 0 . 
\end{align}
This assumption leads, with the help of the variational principle, immediately to an inequality between the
ground-state energies of $H$ and $H_U$, denoted $E_0$ and $\epsilon_0$. 
We have
\begin{align}
\epsilon_0 & = 
\bra{u_0} H_0 \ket{u_0} + \bra{u_{0}} U \ket{u_{0}}  \nonumber \\
& \ge \bra{u_{0}} H_0 \ket{u_{0}} + \bra{u_{0}} V \ket{u_{0}}  
= \bra{ u_{0} } H \ket{ u_{0} } \nonumber \\
& \ge E_{0} .\label{ineq-gs}
\end{align}
Here the second line uses the assumption Eq. \refe{ass}, and the third line uses the Rayleigh--Ritz variational principle:
$u_{0}$ is the wrong wave function for the ground state of $H$.

The ground-state wave function of the spherically symmetric potential $\Vb$, which is the spherical average of $V(\r)$,
will be denoted $u_0=u_0(|\r|)$;
it is an $s$ state, since the ground state wave function has no nodes.
We have 
\begin{multline}\label{u0u0}
\bra{u_0} \Delta V \ket{u_0} =
\int \! \Delta V(\r) |u_0(r)|^2 \, d^3r \\
= \int_0^\infty \!\!\! dr \, r^2 |u_0(r)|^2 \! \int \!\! d\Omega \, \Delta V = 0,
\end{multline}
by using Eq. \refe{dv0}. In the inequality \refe{ineq-gs}, we take $u=u_0$, and the assumption of Eq. \refe{ass} is thus satisfied,
with $\Delta V = V - \Vb$ and $\Vb \equiv U$. Therefore, we conclude that the ground state of 
a potential $V(\r)$, denoted $E_0$, is always lower or equal in energy than the ground state of its
isotropic component $\Vb$, denoted $\bar{E}_0$:
\bel{gs-ineq}
\bar{E}_0 \ge E_0 .
\ee

A consequence of this result concerns the existence of bound states. 
The inequality \refe{gs-ineq}, and the fact that it has been derived variationally, shows that if the angle-averaged potential $\Vb$ has a bound state, the potential $V(\vec{r})$ has a bound state as well.

The ground-state inequality \refe{gs-ineq} can be phrased concisely in terms of perturbation theory. We have
$H = H_U + \Delta V$, where $H_U$ is as before the Hamiltonian with the angle-averaged potential. 
Consider $H_U$ the unperturbed part and $\Delta V$ as a perturbation. The energy shift of the 
ground state in first-order perturbation theory is 
$
\Delta E^{(1)} = \bra{u_0} \Delta V \ket{u_0} = \bra{u_0} V - U \ket{u_0}.
$
Therefore:
if the ground-state energy decreases or stays constant in first-order perturbation theory, then the exact ground-state
energy of the full Hamiltonian is lower or equal than that of the unperturbed Hamiltonian.

\vspace*{-1em}
\section{First excited state}
\label{first-excited}

Now we want to derive an inequality which relates the energy of the first excited state $E_1$ of a potential $V(\r)$ to the energy of the first excited state $\bar{E}_1$ of its spherical average $\Vb(r)$.

We will use the Hylleraas--Undheim variational method, which may be stated as follows.
Let $\phi_1, \ldots, \phi_k$ be any $k$ normalized and mutually orthogonal functions. Consider the
$k \times k$ matrix $H'_{ij}=\bra{\phi_i} H \ket{\phi_j}$; its eigenvalues be $E'_1 \le E'_2 \le \ldots \le E'_k$.
Then we have (assuming that $H$ has at least $k$ bound states)
\bel{HU}
E_i \le E'_i
\qquad
\text{for all } i=1\ldots k .
\ee

The wave functions of the spherically averaged potential $\Vb$ will be denoted $u_0$, $u_1$ for the ground state and the first excited state, respectively.
We take $k=2$, the orthonormal functions $\phi_1=u_0$, $ \phi_2=u_1$;
the matrix $H'$ is 
\bel{HU-M}
H' = 
\begin{pmatrix}
\bar{E}_0 & \bra{u_0} \Delta V \ket{u_1} \\
\bra{u_1} \Delta V \ket{u_0} & \bar{E}_1 + \bra{u_1} \Delta V \ket{u_1}
\end{pmatrix}
\ee
where we have used $H=H_0 + \Vb + \Delta V$ and $\bra{u_0} \Delta V \ket{u_0}$=0 (cf. Eq. \refe{u0u0}).

There are two cases regarding the symmetry of the wave function of the first excited state in a spherically symmetric potential. 
Most often, it is a $p$ state (e.g.~in a deep spherical well or in the three-dimensional anisotropic harmonic oscillator); 
however, there are also potentials where the first excited state is an $s$ state.

Let us first consider the case where the first excited state of $\Vb$ is a $p$ state.
There are three energetically degenerate wave functions $f_1, f_2, f_3$ of the form
$\chi(r) Y_{1 m}(\theta,\phi)$ with $m=-1, 0 ,1$. 
Let us put
$
u_{1}(\vec{r}) = a_1 f_1(\vec{r}) + a_2 f_2(\vec{r})  + a_3 f_3(\vec{r}) 
$
where $\vec{a}=(a_1,a_2,a_3)$ with $|\vec{a}|=1$ is to be determined later.
Consider the quadratic form in the variables $a_1,a_2,a_3$:
\bel{1p-mat}
\bra{u_1} \Delta V \ket{u_1} = \vec{a}^\dag M \vec{a}
\ee
with the hermitian $3 \times 3$ matrix 
$M_{ij} = \bra{f_i}\Delta V \ket{f_j}$. 
One sees that the trace of the matrix $M$ is zero:
\begin{multline}\label{1p-tl}
\text{tr }M = \sum_{i=1}^3 \bra{f_i}\Delta V \ket{f_i} = \\
\int_0^\infty \!\!\! dr \, r^2 |\chi(r)|^2 \!\!\int \!\! d\Omega \, \Delta V \! \left( |Y_{10}|^2 + |Y_{11}|^2 + |Y_{1,-1}|^2 \right) \\
= \frac{3}{4\pi} \int dr r^2 |\chi(r)|^2 \int \! d\Omega \,\Delta V = 0
\end{multline}
where we have used Eq. \refe{dv0} and the identity 
$
\sum_{m=-\ell}^\ell Y^\ast_{\ell m}(\theta,\phi) Y_{\ell m}(\theta,\phi) = \frac{2\ell +1}{4\pi}.
$
The minimal value of the quadratic form in Eq. \refe{1p-mat}, for to any $\vec{a}$ with $|\vec{a}|=1$, is the smallest eigenvalue of $M$, 
which is $\le 0$ here since the trace is the sum of eigenvalues. Therefore there is a linear combination of the three $p$ states, such that
\bel{u1-ass}
\bra{u_1} \Delta V \ket{u_1} \le 0.
\ee

Now we want to conclude $\bra{u_0}\Delta V\ket{u_1}=0$,
which can be often done based on symmetry. As a concrete example, assume that the anisotropic potential $V(\r)$ is symmetric
under inversion, i.e.\ $V(\r) = V(-\r)$;
then, this is also true of $\Delta V$. Thus $\bra{u_0}\Delta V\ket{u_1}=0$ because the $p$ state $u_1$ is odd under inversion.
This argument can be generalized if $V(\r)$ has a ``sufficiently high'' symmetry, this is discussed in Sec.~\ref{sec:symm}.

Applying the Hylleraas--Undheim inequalities \refe{HU} to the matrix in Eq. \refe{HU-M} and using Eq. \refe{u1-ass}, we get the inequality for the first excited state:
\bel{ineq-p}
E_1 \le \bar{E}_1, 
\ee
provided there is sufficient symmetry to have $\bra{u_0}\Delta V\ket{u_1}=0$. 

The case when the first excited state of the isotropic potential $\Vb$ is an $s$-wave state is simpler. Here we have
$\bra{u_0}\Delta V\ket{u_1}=0$
and $\bra{u_1} \Delta V \ket{u_1} = 0$, by the same reasoning as in Eq. \refe{u0u0}, and thus the inequality \refe{ineq-p} follows. 

We have found%
\footnote{%
One can arrive the inequality \refe{ineq-p} using the Rayleigh--Ritz principle for excited states, but
a more complicated discussion of orthogonality would be needed.
}%
, therefore, an inequality for the energy of the first excited state $E_1$
of a potential and the energy of the first excited state $\bar{E}_1$  of its angle-average. It is valid if the potential is symmetric on $\vec{r} \to -\vec{r}$, or has some other ``high'' symmetry (see Sec.~\ref{sec:symm}).

Let us connect the results with perturbation theory, as was done for the ground state. Write $H= H_0 + \Vb + \Delta V$ as before, the isotropic part $H_0+\Vb$ is the unperturbed potential, and $\Delta V$ is considered as a perturbation. 
When the first excited states of the $\Vb$ are $p$ states one needs to apply degenerate perturbation theory.
The energy shifts in first order~\cite{LL3} are precisely the eigenvalues of the matrix $M$ which appears in Eq. \refe{1p-mat},
and it was shown that at least one of the eigenvalues is $\le 0$. The second-order energy shift for the first excited state is 
\begin{multline}
\!\! \Delta E^{(2)}_1 = \sum_{m \ne 1}  \frac{|\bra{u_m} \Delta V \ket{u_1}|^2}{\bar{E}_1 -  
\raisebox{0pt}[1.2\height][\depth]{ $\bar{E}_m$ }
} 
= \frac{|\bra{u_0} \Delta V \ket{u_1}|^2}{\bar{E}_1 - 
\raisebox{0pt}[1.2\height][\depth]{ $\bar{E}_0$ }
} \\ 
+ \sum_{m \ge 2} \frac{|\bra{u_m} \Delta V \ket{u_1}|^2}{\bar{E}_1 - 
\raisebox{0pt}[1.2\height][\depth]{ $\bar{E}_m$ }
} \le 0,
\end{multline}
provided $\bra{u_0} \Delta V \ket{u_1}=0$. 
The case when the first excited state is an $s$ state leads to the same results. 
Thus, the inequality \refe{ineq-p} is valid in first-order perturbation theory regardless of symmetry, and in second-order perturbation theory if $\Delta V$ satisfies a symmetry requirement.

\newpage
\section{Other dimensionalities}
The preceding discussion also holds for the two-dimensional Schr\"odinger equation and its bound states.
One only has to replace the definition of the spherical average, Eq. \refe{sph-av}, by the angle-average in two dimensions
\bel{sph-av-2D}
\bar{V}(\rho) = \frac{1}{2\pi} \int_0^{2\pi} \! V(\rho,\phi) \, d\phi
\ee
where $(\rho,\phi)$ are polar coordinates. 
The inequality for ground states \refe{gs-ineq} holds. One can also repeat the calculations for the $p$ 
excited state, obtaining the inequality \refe{ineq-p}, again
subject to sufficient symmetry (see Sec.~\ref{sec:symm}) to conclude $\bra{u_0} \Delta V \ket{u_1}=0$.

In the one-dimensional case averaging over all space directions becomes the average over two points $x$ and $-x$. 
Thus here the decomposition $V=\Vb + \Delta V$ is the splitting of the potential into its even and odd component.
The ground-state inequality \refe{gs-ineq} remains true. 
It seems that one cannot repeat the reasoning of Sec. \ref{first-excited}, however.

\section{Symmetry considerations}
\label{sec:symm}

Here we discuss necessary conditions to have
\be
\bra{u_0} V\ket{u_1}=
\int \!\! u_0^\ast(\r) V(\r) u_1(\r) d^3r =
0,
\ee
where $u_0, u_1$ are of $s, p$ symmetry, respectively,
and $V(\r)$ is some anisotropic potential. We will conclude that $\bra{u_0} V\ket{u_1}=0$ if $V(\r)$ has ``sufficiently high'' symmetry.

Often a potential $V(\r)$ is symmetric with respect to some geometric transformations, such as rotations or reflections. 
Thus we consider a potential $V$ which is invariant under a point group $G$. 
The wave function $u_0$, being an $s$ state, is also invariant under $G$.
The wave function $u_1$ is a linear combination of the $\ell=1$ spherical harmonics. 
Linear combinations of spherical harmonics (with fixed $\ell$) transform according to one of the irreducible representations of the point group $G$. The general selection rule for a matrix element \cite{LL3}
then tells that $\bra{u_0} V\ket{u_1}=0$ if the $p$ states transform according to a non-trivial representation of $G$. Note that we need to look at all $p$ states since in Sec. \ref{first-excited} we have constructed $u_1$ to be a linear combination of all three states.

When the point symmetry group $G$ is known, one can look up in group tables how the spherical harmonics of $p$ type 
transform under this group. For the convenience of the reader, we state the results here.
One has $\bra{u_0} V\ket{u_1}=0$ if
$G$ is any of the following groups:
i) the seven cubic and icosahedral groups,
ii) the axial groups $C_{nh}$, $D_n$, $D_{nh}$, $D_{nd}$, $S_{2n}$ for $n$ integer, {\em except} $C_{1h}\!=\!C_s$ and $D_{1h}\!=\!C_{2v}$,
iii) the cylindrical group $D_{\infty h}$.
The case from Sec. \ref{first-excited}, where $\bra{u_0} V\ket{u_1}=0$ if $V(\r)$ is invariant under 
inversion,
corresponds to symmetry group $S_2$. 


In the two-dimensional case the same arguments apply, only that we are dealing with a two-dimensional point group $G$. We have $\bra{u_0} V\ket{u_1}=0$ if $G$ is one of the two-dimensional point groups $C_n^{(2d)}$, $D_n^{(2d)}$ for $n \ge 2$. For example, invariance under inversion (same as rotation by $\pi$) is sufficient, this is symmetry $C_2^{(2d)}$.

\section{Conclusions}

We have discussed how bound-state energies of an arbitrary anisotropic potential and its spherical average are related. An exact inequality for the ground states and for the first excited states was given. While the ground-state result is general, the excited-state inequality requires that the anisotropic potential possesses certain symmetry. Connections with perturbation theory are discussed, and a remark on the existence of bound states is made. The results apply also in the two-dimensional case.

The results may be useful for the practical calculation of energies of bound states. 
It is relatively easy to calculate bound-state energies for spherically symmetric potentials, while full the numerical solution of a three-dimensional problem can be difficult.
Bounds, such as those discussed here, provide an estimate for the eigenvalues of the three-dimensional problem.


\begin{thebibliography}{99}

\bibitem{Pais-Uhlenbeck}
A. Pais and G. E. Uhlenbeck,
{\em On the quantum theory of the third virial coefficient},
Phys. Rev. {\bf 116}, 250 (1959).

\bibitem{Klawunn-Recati}
M. Klawunn and A. Recati,
{\em Fermi polaron in two dimensions: Importance of the two-body bound state},
Phys. Rev. A {\bf 84}, 033607 (2011).

\bibitem{LL3}
L. D. Landau and E. M. Lifshitz, 
{\em Quantum Mechanics}
(Pergamon Press, Oxford, 1977).

\bibitem{Beesack}
P. R. Beesack, 
{\em A note on an integral inequality}, 
Proc. Am. Math. Soc. {\bf 8}, 875 (1957).

\bibitem{Banks}
D. Banks,
{\em An integral inequality},
Proc. Am. Math. Soc. {\bf 14}, 823 (1963).

\bibitem{Note1}
One can arrive the inequality \refe{ineq-p} using the Rayleigh--Ritz principle for excited states, but
a more complicated discussion of orthogonality would be needed.





\end{thebibliography}
\end{document}